\documentclass[article]{aa501}
\usepackage{amsmath,amssymb}
\usepackage{graphicx}
\usepackage{natbib}
\newcommand{\rmd}{{\rm d}}
\newcommand{\bl}{{\mathbf{l}}}
\newcommand{\bL}{{\mathbf{L}}}
\newcommand{\br}{{\mathbf{r}}}
\newcommand{\bx}{{\mathbf{x}}}

\newcommand{\BR}{{\mathbb{R}}}
\newcommand{\CA}{{\cal A}}
\newcommand{\CF}{{\cal F}}
\newcommand{\CM}{{\cal M}}

\newcommand{\hMpc}{{\ifmmode{h^{-1}{\rm Mpc}}\else{$h^{-1}$Mpc}\fi}}
\newcommand{\hMsun}{{\ifmmode{h^{-1}M_\odot}\else{$h^{-1}M_\odot$}\fi}}
\newcommand{\paverage}[1]{\left\langle #1 \right\rangle_{\rm P}}
\newcommand{\cov}{{\ifmmode{\text{{\em cov}}}\else{ {\em cov} }\fi}}
\newcommand{\var}{{\ifmmode{\text{{\em var}}}\else{ {\em var} }\fi}}
\begin{document}

\title{Correlations in the Orientations of Galaxy Clusters}

\author{
A.~Faltenbacher\inst{1} \and
S.~Gottl\"ober\inst{1} \and
M.~Kerscher\inst{2} \and 
V.~M\"uller\inst{1}
}
\institute{
Astrophysikalisches Institut Potsdam, 
An der Sternwarte 16
14482 Potsdam, Germany
\and
Sektion Physik, Ludwig-Maximilians-Universit{\"a}t, 
Theresienstra{\ss}e 37, 80333 M{\"u}nchen, Germany
} 
\titlerunning{Correlation in the Orientations of Galaxy Clusters}
\authorrunning{Faltenbacher et al.}

\abstract{ The relative orientation of clusters' major elongation axes
and clusters' angular momenta is studied using a large N-body
simulation in a box of 500~\hMpc\ base length for a standard
$\Lambda$CDM model.  Employing the technique of mark correlation
functions, we successfully separated the correlations in the
orientation from the well known clustering signal traced by the
two-point correlation function.  The correlations in the orientation
are highly significant for our sample of 3000 clusters.  We found an
alignment of neighboring clusters, i.e. an enhanced probability of
the major elongation axes of neighboring cluster pairs to be in parallel
with each other. At 10~\hMpc\ separation the amplitude of this signal is
$\sim 10\%$ above the value expected from random orientations, and it
vanishes on scales larger than 15~\hMpc. The ``filamentary'' alignment
between cluster's major elongation axes and the lines pointing towards
neighboring clusters shows even stronger deviations from random
orientation, which can be detected out to scales of 100~\hMpc, both in
2D and 3D analyses. Similarly, strong correlations of the angular
momentum were seen. 
Also a clear signal in the scalar correlation of the absolute value of
the angular momentum, the spin parameter and the mass was found.  They
extend up to 50~\hMpc\ and have an amplitude of 40\%, 15\%, and 10\%
above a random distribution at 10~\hMpc\ separation, respectively.
}
\maketitle
\section{Introduction}
The study of orientation effects between galaxy clusters has a long
and controversial history in cosmology.  In a seminal study
{}\citet{binggeli:shape} claimed that galaxy clusters are highly
eccentric and oriented relative to neighboring clusters if lying at
separations smaller than 15~\hMpc.  Further he found anisotropies in
the cluster distribution on scales up to 50~\hMpc.  Following studies
found no or weak statistical significance for orientation effects
between neighboring clusters or between cluster orientation and the
orientation of the central dominant galaxy, cp.
{}\citet{struble:new}, {}\citet{flin:alignment}, and 
{}\citet{rhee:mirror}. Remarkable was the apparent absence
{}\citep{ulmer:major} or weakness {}\citep{rhee:x+opt} of orientation
effects in projected X-ray contours of clusters, but it should be
noted that the cluster samples at this time were small.  Analyzing a
large set of 637 Abell clusters, {}\citet{plionis:up150} found highly
significant alignment effects on scales below 10~\hMpc\ that become
weaker but extend up to 150~\hMpc. More objectively selected,
but smaller cluster samples seemed to put into question the reality of
this signal, cp.  {}\citet{fong:2d} and
{}\citet{martin:milano}. However, {}\cite{chambers:x-contours} found
significant nearest neighbor alignment of cluster X-ray isophotes
using data from {\sl Einstein} and {\sl ROSAT}.  With the advent of
new rich cluster catalogues as the optical {}\textsc{Enacs} survey
{}\citep{katgert:enacs-i} and the X-ray based {}\textsc{Reflex} survey
{}\citep{boehringer:reflex}, the question of orientation effects in
clusters should attract renewed attention. Sufficiently large and well
defined cluster samples showing only weak contamination by projection
effects seem to be necessary to clarify this uncertain situation.  

Strong stimulus to study orientation effects in clusters came from
early ideas that a possible relative orientation between neighboring
clusters or of clusters in the same supercluster should reflect the
underlying structure formation mechanism.  \cite{binney:prolat}
proposed that tidal interactions of evolving protocluster systems may
lead to the growth of anisotropies of clusters and to relative
orientation effects.  Later, \cite{vanHaarlem:merging} used numerical
simulations of CDM models to demonstrate that clusters are elongated
along the incoming direction of the last major merger.  In the same spirit,
\cite{west:merging} found that clusters grow by accretion and merging
of surrounding matter that falls into the deep cluster potential wells
along sheet-like and filamentary high density regions.  Therefore, the
cluster formation is tightly connected with the supercluster network
that characterizes the large-scale matter distribution in the
universe.  High-resolution simulations showing this effect are
described by the Virgo collaboration, cp.  {}\citet{colberg:virgo}.
{}\citet{onuora:alignment} found a significant alignment signal up to
scales of 30~\hMpc\ for a $\Lambda$CDM model, whereas in a 
$\tau$CDM model the signal extended only up to scales of
15\hMpc. 

To quantify the alignment of the galaxy clusters, we use a large
$\Lambda$CDM simulation in a box of 500~\hMpc\ side length. We
identify a set of 3000 clusters. As statistical tools we employ mark
correlation functions (MCF), as introduced to cosmology by
{}\citet{beisbart:luminosity}.  In this article we will extend this
formalism to allow for vector valued marks.  The direction of the
major axis of the mass ellipsoid serves as the vector mark.  Tightly
connected with the elongation of clusters is its internal rotation.
According to {}\cite{doroshkevich:origin} and {}\cite{white:angular},
the primary angular momentum of bound objects is due to tidal
interaction between the elongated protostructures after decoupling
from cosmic expansion and before turn-around.  More recent studies
find that the angular momentum of dark matter halos is later modified
by the merging history of their building blocks, cp.
{}\cite{vitvitska:origin} and {}\cite{porciani:testingI,
porciani:testingII}.  Therefore, we utilise the angular momentum as
an additional mark for the study of the correlation of inner
properties of simulated clusters, and we compare it with the
orientation effects. 

The plan of the paper is as follows.  In the next section, we describe
our numerical simulation, the selection of a cluster sample and the
precision with which we can determine structure parameters from it.
Next we discuss the MCFs that are relevant for our studies.  In
particular, we use special MCFs for vector marks to quantify
correlations of orientation.  In Sect.~\ref{sect:shape} we 
investigate correlations in the spatial orientation of clusters
both in 3D and in the projected mass distribution.  In
Sect.~\ref{sect:angular} we present a MCF analysis using the angular
momentum, mass and spin taken as vector and scalar marks,
respectively. We conclude with a summary of the results.

\section{Cluster sample in numerical simulations}
\label{sect:simulation}
We utilize the AP3M code of \citet{couchman:mesh} to follow the
dynamics of $256^3$ particles in a box of $500~\hMpc$ with periodic
boundary conditions.  We employ a cold dark matter model with a
cosmological constant $\Omega_{\Lambda}=0.7$, a matter density
$\Omega_{\rm m}=1-\Omega_{\Lambda}$, and a Hubble constant $H_0=100 h$
km/s/Mpc with $h=0.7$.  The age of the universe in this model is
$\approx13.5$~Gyrs.  The normalization, given by the linear mass
variance of dark matter on $8~\hMpc$ scale, $\sigma_8=0.87$, is in
accordance with the four year {\sl COBE} DMR observations as well as
the observed abundance of galaxy clusters.  The code uses glass-like
initial conditions, cp.\  \cite{knebe:formation}. The initial power
spectrum was calculated with the CMBFAST code \citep{seljak:cmbfast}.
We start the simulations at an initial redshift $z=25$. Up to this
time the Zeldovich approximation provides accurate results on the
scales considered here.  We employ a comoving softening length of 100
$h^{-1}$ kpc, and 1000 integration steps that are enough to avoid
strong gravitational scattering effects on small scales,
cp. ~\cite{knebe:resolution}.  With this softening length, the 
inner cores or break radii of the cluster sized halos are
resolved. For our statistical analysis the big simulation volume
provides us with a sufficient number of clusters embedded into the
large scale structure network.  In particular an accurate
representation of the large scale tidal field is important for getting
stable orientation results.  The particle mass in the simulation is
$6.2 \times10^{11} \hMsun$, comparable to the mass of one single 
galaxy. For reliable cluster orientations and angular momenta, the
number of particles per cluster should not drop below a minimum of a few
hundred of particles.

We search for bound systems using a friends-of-friends algorithm.  In
our $\Lambda$CDM model the virialization overdensity
$\rho/\rho_{\rm mean}$ at $z=0$ is $\simeq330$
{}\citep{kitayama:semianalytic} which corresponds (for spherical
isothermal systems) to a linking length of 0.17 times the mean inter
particle distance.  These bound systems are our clusters of galaxies,
i.e.\ our cluster sized halos.  The 3000 most massive ones are the
constituents of our mock cluster sample.  The mean distance between
clusters in this sample is $34.7~\hMpc$, which is in agreement with
the mean distance of clusters in the {}\textsc{Reflex} cluster survey
{}\citep{boehringer:reflex}.  The most massive cluster has a mass of
$2.3\times10^{15}\hMsun$, resolved with 3700 particles, the lightest
still has $1.4\times10^{14}\hMsun$ which corresponds to 224 particles.

\begin{figure}
\begin{center}
\resizebox{0.95\hsize}{!}{\includegraphics{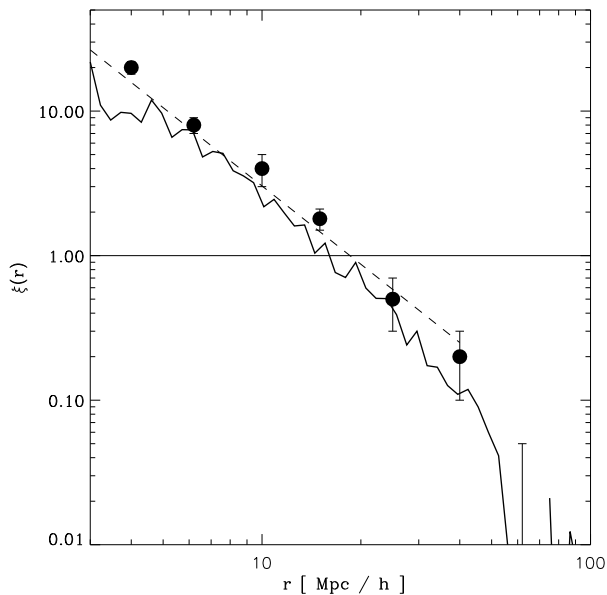}}
\end{center}
\caption{The two-point correlation function of cluster sized halos in
redshift space (solid line).  The dashed line has a slope of
$-1.8$. The data points are taken from \citet{collins:spatial}.
\label{fig:corr} }    
\end{figure}

The two-point correlation function $\xi(r)$ of the cluster sized halos
in redshift space shows the expected behavior (Fig.~\ref{fig:corr})
compatible with the correlation function of the {}\textsc{Reflex}
clusters determined in \citet{collins:spatial}.  The correlation length 
is $16~\hMpc$, quantifying the amplitude of the correlation function. 
At scales of about $70~\hMpc$ $\xi(r)$ becomes negative.

The principal axes of a cluster are determined by means of the 
eigenvectors of the inertial tensor using all halo mass points. 
Assuming an ellipsoidal shape these vectors can be uniquely
transformed into the three axes of the underlying mass ellipsoid.
Subsequently the three axes are assumed to be ordered by size as 
$~a\ge b\ge c~$. 
In addition to the mass ellipsoid, the angular momentum vector $\bL$
and the spin parameter $\lambda$ can be utilized to characterize the 
dynamical state of a cluster sized halo. $\bL$ is calculated by
summing up the angular momenta of all mass particles of the respective 
halo. The spin parameter is defined by $\lambda =
\omega/\omega_{\rm sup}$, where $\omega$ denotes the actual angular
velocity and $\omega_{\rm sup}$ the angular velocity needed for the system
to be rotationally supported against gravity.  This ratio can be
expressed as  
\begin{equation}  
\label{equ:spin}
\lambda = L/(G^{1/2}M^{3/2}R^{1/2}),
\end{equation}  
where $L$ is the absolute value of the angular momentum $\bL$, $M$ the
total mass of the system and $R$ its radius, respectively (e.g.\
{}\citealt{padmanabhan:structure}). The radius $R$ is taken as the
radius of a sphere with the same volume as the halo (estimated on a
fine grid).  
\section{Mark correlation functions}
\label{sect:markcorr}
Mark correlation functions (MCFs) have been used for the
characterization of marked points sets for quite some time (see
\citealt{stoyan:fractals} for an instructive introduction and
overview). {}\citet{beisbart:luminosity} introduced MCFs to cosmology
and used them to quantify the luminosity and morphology dependent
clustering in the observed galaxy distribution (see also
\citealt{szapudi:correlationspscz}).  {}\citet{gottloeber:spatial}
discussed the correlations of galaxy sized halos depending on the
merging history with MCFs (see also {}\cite{gottloeber:merging}).  In
a recent review {}\citet{beisbart:mark} show other physical
applications of the MCFs and also how the MCFs can be calculated from
stochastic models.  To make this article self contained, we repeat
briefly the basics of MCFs and then extend the formalism to allow for
vector marks:

Consider the set of $N$ points $\{\bx_i\}_{i=1}^{N}$ and attach a mark
$m_i$ to each point $\bx_i\in\BR^3$ resulting in the marked point set
$\{(\bx_i,m_i)\}_{i=1}^{N}$
{}\citep{stoyan:oncorrelations,stoyan:fractals}.  In the following we
use a variety of marks: the mass, the absolute value of the angular
momentum, the spin parameter $\lambda$, and as vector valued marks,
the principal axes of the second moment of the mass distribution and
the angular momentum.  Let $\rho$ be the mean number density of the
points in space and $\rho^M(m)$ the probability density of the mark
distribution. The mean mark is then $\overline{m}=\int\rmd m\
\rho^M(m)m$, and the variance is $V[m]=\int\rmd m\ \rho^M(m)
(m-\overline{m})^2$. We assume that the joint probability
$\rho^{SM}(\bx,m)$ of finding a point at position $\bx$ with mark $M$,
splits into a space-independent mark probability and the constant mean
density: $\rho^M(m)\times\rho$.
\newline
The spatial-mark product-density
\begin{equation}
\rho_2^{SM}((\bx_1,m_1),(\bx_2,m_2))\ 
\rmd V_1 \rmd m_1\ \rmd V_2 \rmd m_2 ,
\end{equation}
is the joint probability of finding a point at $\bx_1$ with the mark
$m_1$ and another point at $\bx_2$ with the mark $m_2$.  We obtain the
spatial product density $\rho_2(\bx_1,\bx_2)$ and the two-point
correlation function $\xi(r)$ by marginalizing over the marks:
\begin{multline}
\rho^2\ (1+\xi(r)) = \rho_2(\bx_1,\bx_2) = \\
=\int\rmd m_1 \int\rmd m_2\ \rho_2^{SM}((\bx_1,m_1),(\bx_2,m_2)) ,
\end{multline}
with $\xi(r)$ only depending on the distance $r=|\bx_1-\bx_2|$ of the
points in an homogeneous and isotropic point set.
\newline
We define the conditional mark density:
\begin{multline}
\CM_2(m_1,m_2|\bx_1,\bx_2) =\\
=\begin{cases}
\frac{\rho_2^{SM}((\bx_1,m_1),(\bx_2,m_2))}{\rho_2(\bx_1,\bx_2)}
& \text{ for } \rho_2(\bx_1,\bx_2)\ne 0,\\
0 & \text{ else }.
\end{cases} 
\end{multline}
For a stationary and isotropic point distribution,
$\CM_2(m_1,m_2|r)\rmd m_1\rmd m_2$ is the probability of finding the
marks $m_1$ and $m_2$ of two galaxies located at $\bx_1$ and $\bx_2$,
under the condition that they are separated by $r=|\bx_1-\bx_2|$. Now
the full mark product-density can be written as
\begin{multline}
\rho_2^{SM}((\bx_1,m_1),(\bx_2,m_2)) = \\
=\CM_2(m_1,m_2|\bx_1,\bx_2)\ \rho_2(\bx_1,\bx_2) .
\end{multline}
If there is no mark-segregation $\CM_2(m_1,m_2|r)$ is independent from $r$, and
$\CM_2(m_1,m_2|r)=\rho^M(m_1)\rho^M(m_2)$.

Starting from these definitions, especially using the conditional mark
density $\CM_2(m_1,m_2|r)$, one may construct several mark-correlation
functions sensitive to different aspects of mark-segregation
{}\citep{beisbart:luminosity}.  The basic idea was to consider
weighted correlation functions conditional that two points can be
found at a distance $r$.
For a non-negative weighting function $f(m_1,m_2)$ we define the
average over pairs with separation $r$:
\begin{equation}
\label{eq:def-paverage}
\paverage{f}(r) = \int\rmd m_1\int\rmd m_2\ f(m_1,m_2)\ 
\CM_2(m_1,m_2|r).
\end{equation}
$\paverage{f}(r)$ is the expectation value of the weighting function
$f$ (depending only on the marks), under the condition that we find a
galaxy-pair with separation $r$.  For a suitably defined
integration measure, Eq.~\eqref{eq:def-paverage} is also applicable to
discrete marks.
The definition~\eqref{eq:def-paverage} is very flexible, and allows us
to investigate the correlations both of scalar and vector valued
marks.

To calculate such mark correlation functions from the cluster data we
use an estimator taking into account the periodic boundaries of the
simulation box. We obtain virtually identical results for the
estimator without boundary corrections {}\citep{beisbart:luminosity}.

\subsection{Correlations of scalar marks}
\label{sect:mcf-scalar}
For scalar marks the following mark correlation functions have proven
to be useful {}\citep{beisbart:luminosity}: 
\begin{itemize}
\item
The simplest weight to be used is the mean mark, 
\begin{equation}
k_{\rm m}(r) \equiv \frac{\paverage{m_1+m_2}(r)}{2\ \overline{m}} .
\end{equation}
that quantifies the deviation of the mean mark of pairs with separation $r$
from the overall mean mark $\overline{m}$.  A $k_{\rm m}>1$ indicates mark
segregation for point pairs with a separation $r$, specifically their
mean mark is larger than the overall mark average.\\
\item
Accordingly, higher moments of the marks may be used to quantify mark
segregation. The mark variance $\var(r)$ 
\begin{equation}
  \label{eq:def-var}
  \var(r) \equiv \paverage{\left(m_1-\paverage{m_1}(r)\right)^2}(r). 
\end{equation}
gives information about the fluctuations of the marks of points
which are separated by a distance $r$ to other members of the set.  
A $\var(r)/V[m]$ larger than one indicates a substantially increased
scattering of the marks compared to the overall variance.
\item 
The mark covariance {}\citep{cressie:statistics} is
\begin{align}
\label{eq:def-cov}
\cov(r) & \equiv \paverage{ \big(m_1-\paverage{m_1}(r)\big)\big(m_2-\paverage{m_2}(r)\big)}(r)
\nonumber \\
        & =  \paverage{m_1 m_2}(r) - \paverage{m_1}(r)\paverage{m_2}(r).
\end{align}
Mark segregation can be detected by looking whether $\cov(r)$ differs
from zero. A $\cov(r)$ larger than zero, e.g., indicates that points
with separation $r$ tend to have similar marks.  
\end{itemize}
\subsection{Correlations of vector marks}
To study the alignment effects between the clusters we attach to each
cluster the orientation as a vector mark $\bl$ with $|\bl|=1$. The
orientation is either given by the direction of the major half-axis of
the mass ellipsoid, or the direction of the angular momentum.  The
distance vector between two clusters is $\br$, and the normalized
direction is $\hat\br=\br/r$. We consider the following MCFs:
\begin{equation}
\label{eq:def-vector-corr}
\begin{split}
\CA(r) &= \paverage{|\bl_1\cdot\bl_2|}(r), \\
\CF(r) &= \frac{1}{2}
\paverage{|\bl_1\cdot{\hat\br}| + |\bl_2\cdot{\hat\br}|}(r), 
\end{split}
\end{equation}
where ``$\cdot$'' denotes the scalar product. Due to the spatial
symmetry of the mass ellipsoids we use the absolute values of these
scalar products.  
\begin{itemize}
\item
$\CA(r)$ quantifies the direct $\CA$lignment of the vectors $\bl_1$
and $\bl_2$ (see also \citealt{stoyan:fractals} and their $k_d$).
$\CA(r)$ is proportional to the cosine of the angle between $\bl_1$
and $\bl_2$.
\item
$\CF(r)$ quantifies the $\CF$ilamentary alignment of the vectors
$\bl_1$ and $\bl_2$ with the line connecting both clusters. $\CF(r)$
is proportional to the cosine of the angle between $\bl_1$ and the
direction vector $\hat\br$ connecting the points.
\end{itemize}
In three dimensions no mark segregation implies
$\CA(r)=\CF(r)=0.5$, in two dimensions $\CA(r)=\CF(r)=2/\pi$.
{}\citet{beisbart:mark} provide some further explanations and discuss
in which sense $\CA(r)$, $\CF(r)$ provide a complete characterization
of the vector correlations.

\section{Correlations in the orientation of clusters' shape}
\label{sect:shape}
\begin{figure}
\begin{center}
\resizebox{0.7\hsize}{!}{\includegraphics{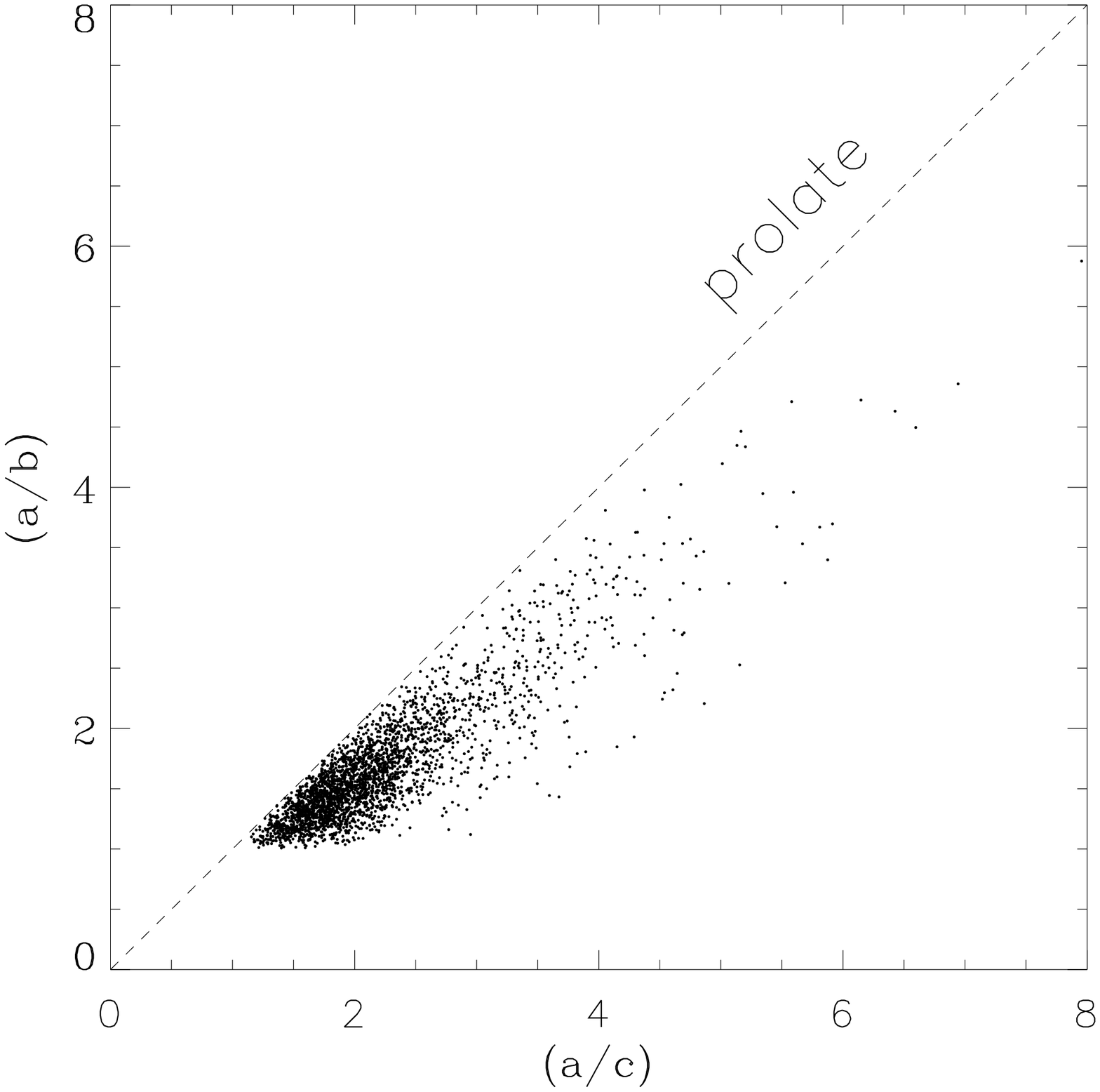}}
\resizebox{0.7\hsize}{!}{\includegraphics{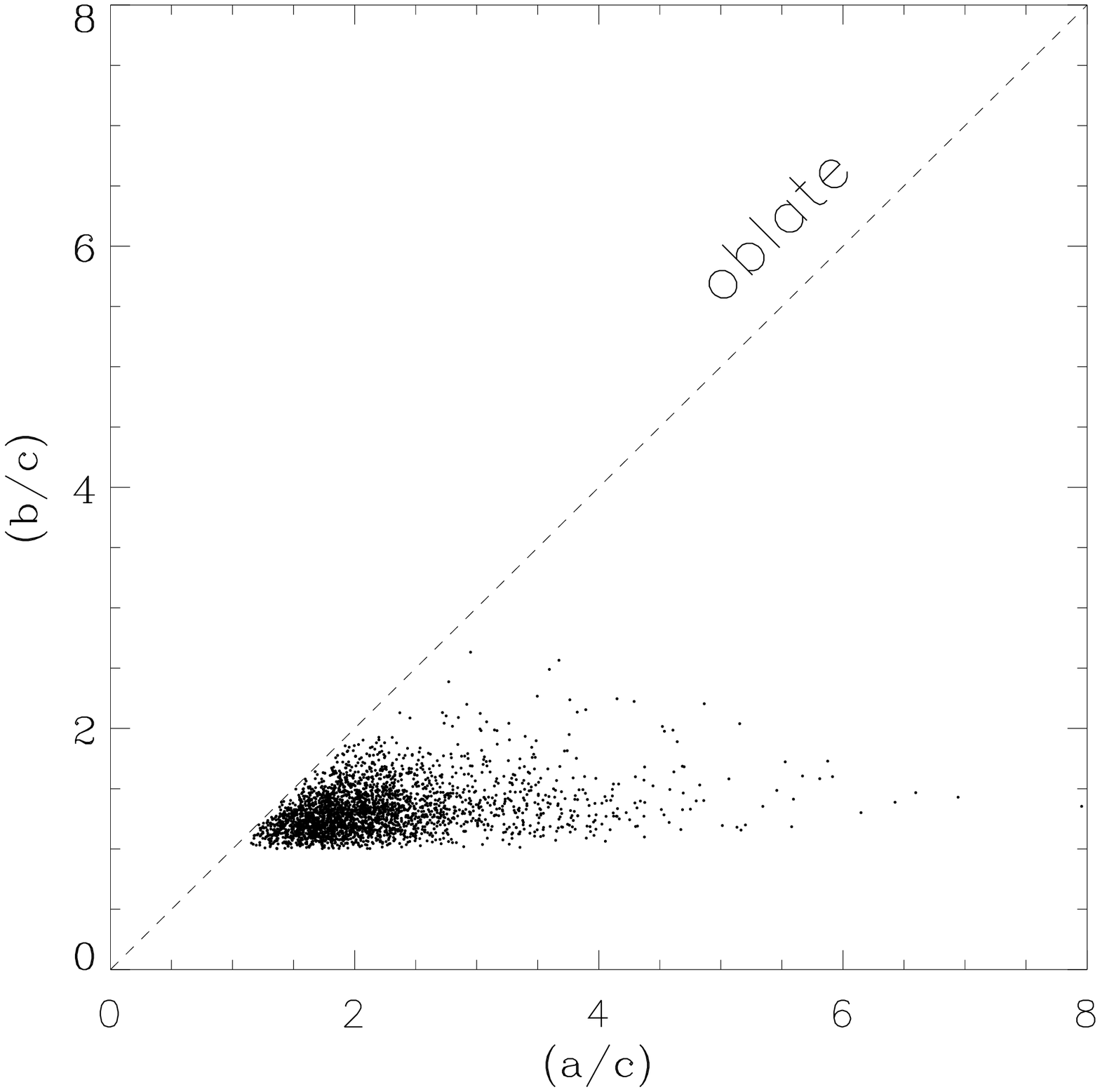}}
\end{center}
\caption{Ratios of the length of the clusters principal axes 
($a\ge b\ge c$). Upper panel: $a/b$ versus $a/c$, for {\em prolate}
rotational ellipsoids ($a>b$, $b = c$) all points would coincide with
the dashed diagonal; lower panel: $b/c$ versus $a/c$, for {\em oblate}
rotational ellipsoids ($a=b$, $a>c$) all data points would coincide
with the dashed diagonal.   
 \label{fig:proobl}}
\end{figure}

As described in Sect.~\ref{sect:simulation} we determine the mass
ellipsoid for each of the cluster sized halos. The distribution of the
intrinsic shapes of clusters in our sample is illustrated in
Fig.~\ref{fig:proobl}. There we plot the ratios between the lengths of
the three principal axes $a\ge b\ge c$: the upper panel shows the
ratios $a/b$ versus $a/c$, whereas the lower panel shows $b/c$ versus
$a/c$.  For prolate rotational ellipsoids ($a>b$, $b = c$) all points
would coincide with the dashed diagonal on the upper panel.  In the
case of oblate rotational ellipsoids ($a=b$, $a>c$) all data points
would coincide with the dashed diagonal plotted in the lower panel.
Clearly the majority of the data points tend to be concentrated rather
along the upper panels' diagonal than along the lower panels'
diagonal.  Therefore, the clusters in our simulations typically have
prolate shape, in good agreement with observations (e.g.
{}\citealt{cooray:prolate}) and other CDM simulation, cp.\
{}\citep{cole:haloes}.

\begin{figure}
\begin{center}
\resizebox{0.7\hsize}{!}{\includegraphics{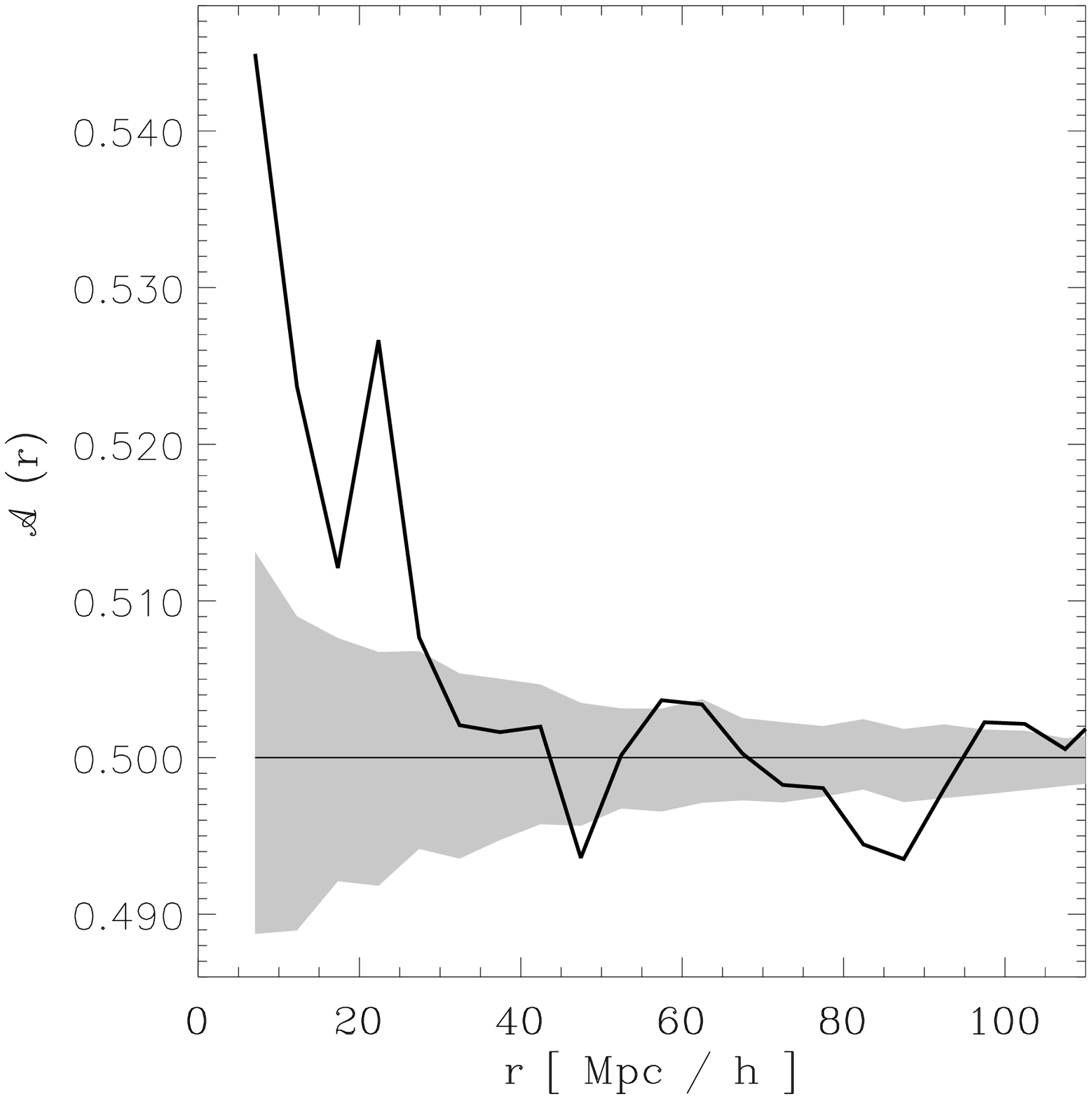}}
\resizebox{0.7\hsize}{!}{\includegraphics{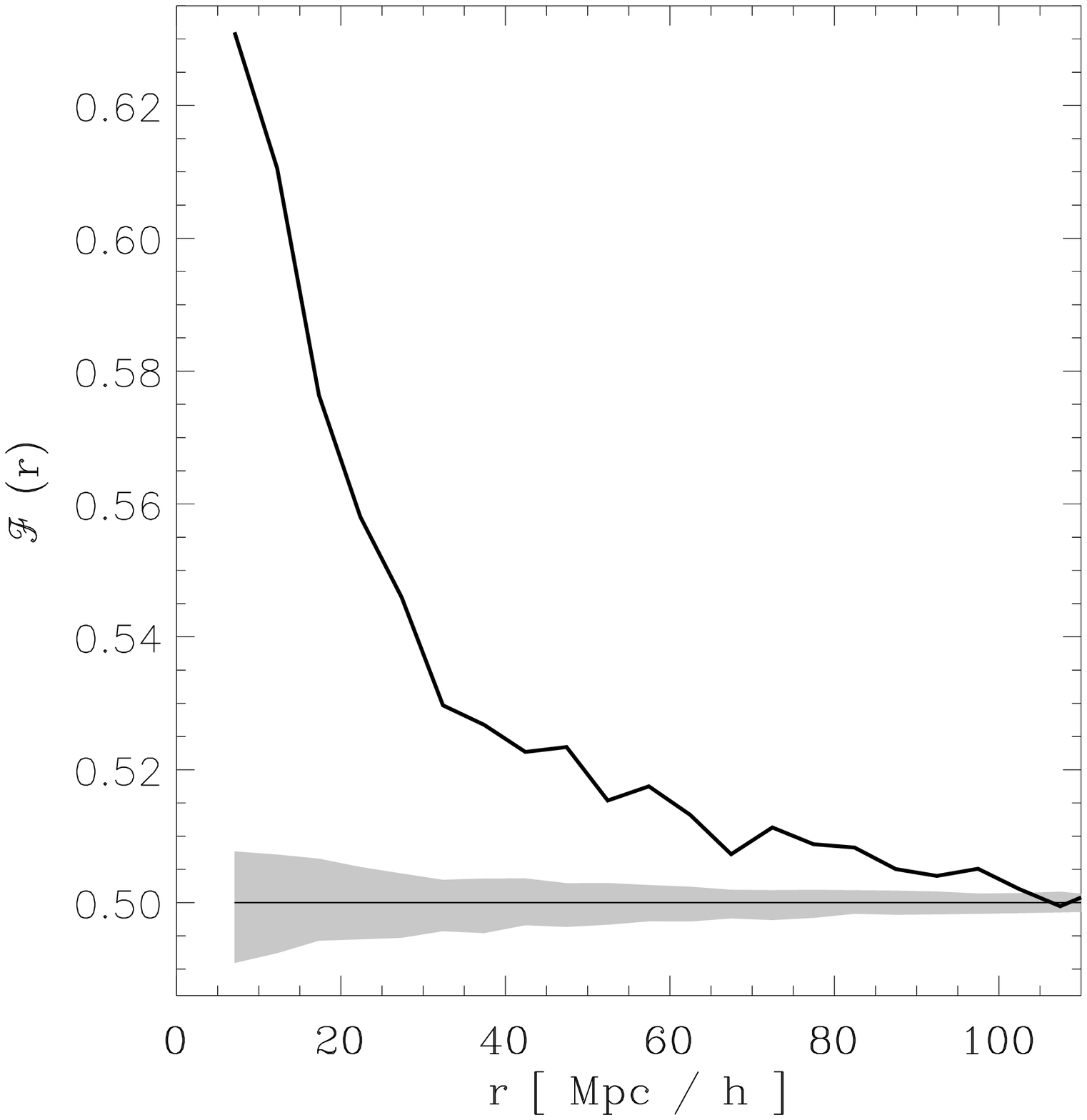}}
\end{center}
\caption{Mark correlations using the {\em 3D} orientation of the dark
  matter halo, specified by the direction of the major axis $\bl$ of
  the mass ellipsoid, as vector mark. The shaded area is obtained by
  randomizing the orientation among the clusters.
\label{fig:clusters-500-orientation}}    
\end{figure}

\begin{figure}
\begin{center}
\resizebox{\hsize}{!}{\includegraphics{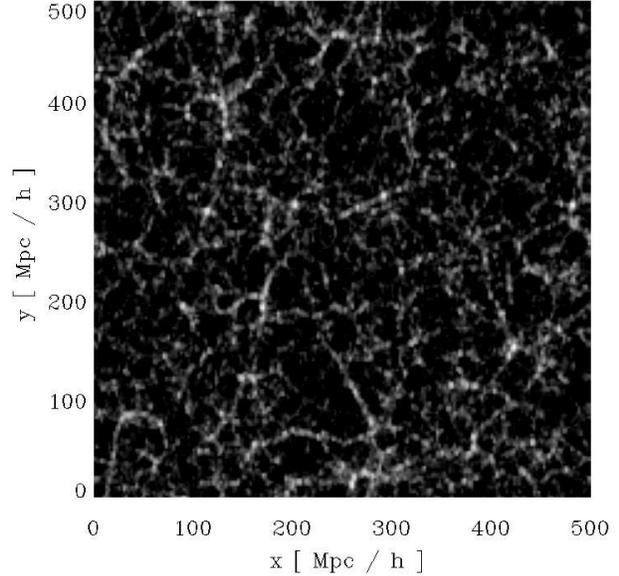}}
\end{center}
\caption{ Projected density field of a $\sim7.5~\hMpc$ slice through 
  the simulation box. The density is calculated on a $300^3$
  grid which corresponds to a smoothing length of $\sim1.7~\hMpc$. The
  maximum value of projected density is $\approx 1.5\times 10^4 
  \rho/\rho_{\rm mean}$. Gray scales are chosen according to $\log$ of
  density.  
\label{fig:densityfield}}
\end{figure}

In the following we will use the direction of the major axis of the
mass ellipsoid (the ``$a$-axis'') as a vector mark
$\bl$. Fig.~\ref{fig:clusters-500-orientation} shows the mark
correlation functions $\CA(r)$ and $\CF(r)$. 
The increased $\CA(r)$ towards smaller scales indicates that pairs of
clusters with a distance smaller than 30~\hMpc\ prefer a parallel
orientation of their orientation axes $\bl_1$ and $\bl_2$. This
deviation of $\sim 10\%$ from the signal of a purely random orientation
is clearly outside the random fluctuations shown by the shaded region.
We determine these random fluctuations by repeatedly and randomly
redistributing the orientation axes among the clusters. The positions
of the clusters remain fixed.
The filamentary alignment of the orientation of the clusters $\bl_1$
and $\bl_2$ towards the connecting vector $\hat\br$ as quantified by
$\CF(r)$ is significantly stronger (up to 20\% deviation from the
signal of purely random orientation on small scales). Remarkably this
signal extends out to 100~\hMpc. 
In a qualitative picture this may be explained by a large number of
clusters elongated in the direction of the filaments of the large
scale structures. Filaments are prominent features found in
the observed galaxy distribution {}\citep{huchra:cfa2s1} as well as in
$N$-body simulations {}\citep{melott:generation}, often with an extent
up to 100~\hMpc.  The inspection of density field of the simulation
utilized here confirms this view (see Fig.~\ref{fig:densityfield}).

\begin{figure}
\begin{center}
\resizebox{0.7\hsize}{!}{\includegraphics{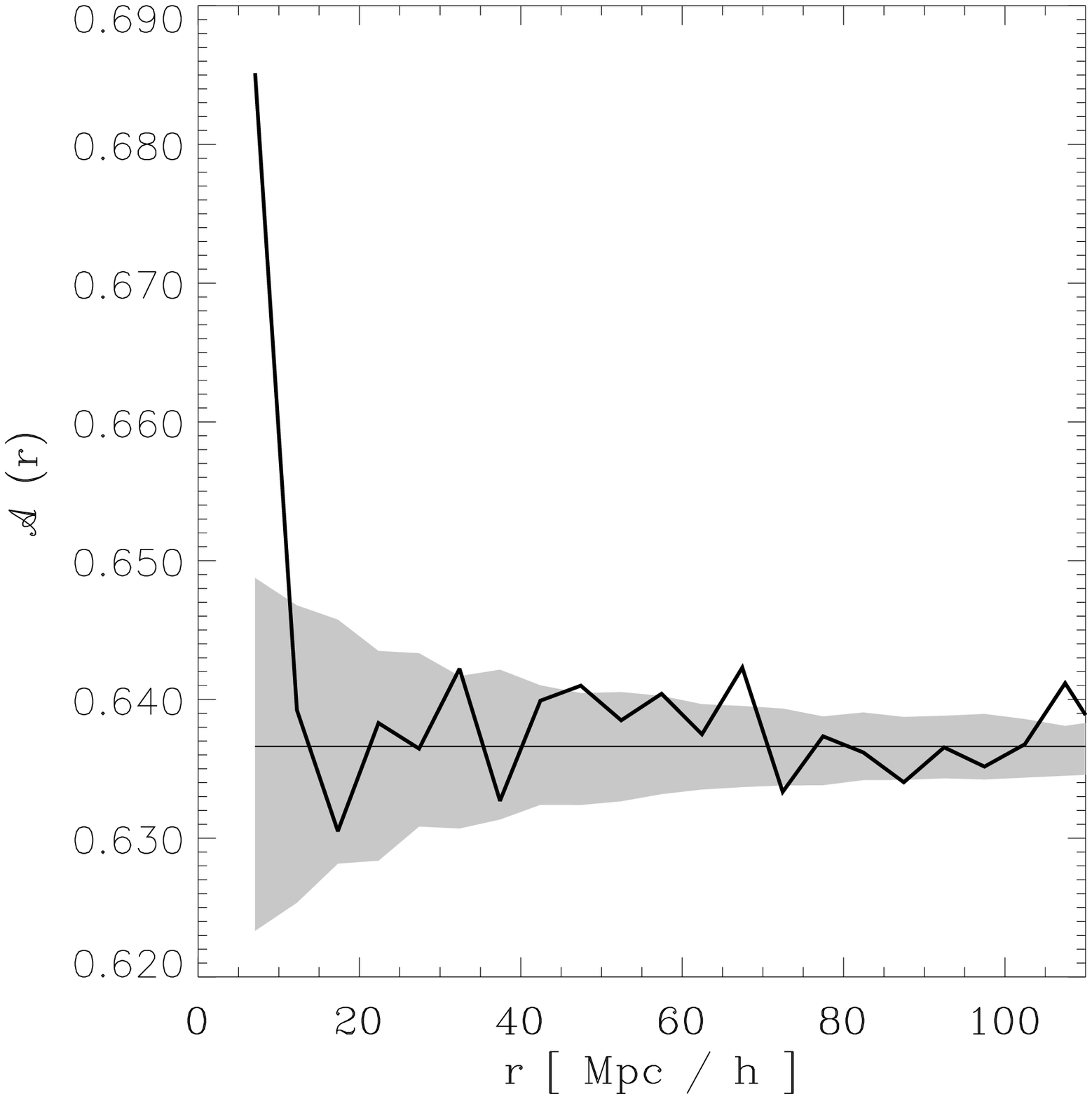}}
\resizebox{0.7\hsize}{!}{\includegraphics{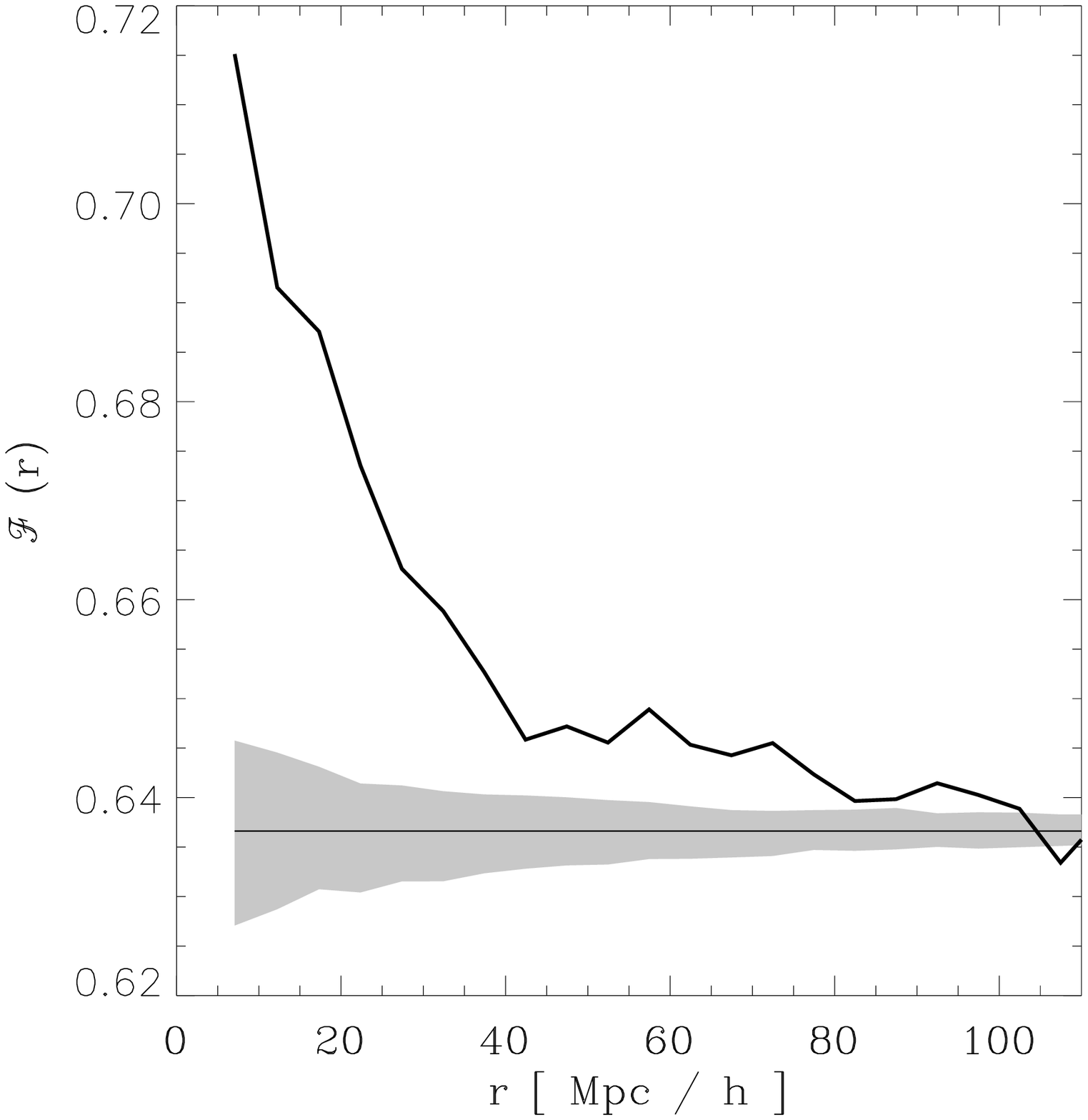}}
\end{center}
\caption{Mark correlations with the normalized major axis $|\bl|=1$ of
  the projected cluster-density ({\em 2D} orientation) as vector
  mark. The shaded area is obtained by randomizing the orientation
  among the clusters. 
\label{fig:clusters-500-proj-orient}}
\end{figure}

Up to now we used the orientation of the mass ellipsoid in three
dimensional space. Most observations, however, only provide the
projected galaxy number density or X-ray intensity. In the following
we will investigate whether the clear signal found in
Fig.~\ref{fig:clusters-500-orientation} will be reduced by projection.
In close analogy to our three dimensional analysis we determine the
orientation of the mass-ellipse of a halo from the mass-density orthogonally
projected onto a side of the simulation box.  We repeat our analysis
with the major axis of the mass ellipse in two-dimensions as a
two-dimensional vector mark. Also the normalized direction $\hat\br$
is given in the plane. However, for the radial distance $r$ we use the
three dimensional separation of the clusters. With this setting, we
mimic the observational constraints, e.g. in the {}\textsc{Reflex}
cluster survey.
In this projected sample the signal in $\CA(r)$ is only visible on
scales smaller than 10~\hMpc\
(Fig.~\ref{fig:clusters-500-proj-orient}).  The amplitude of the
deviation from random orientation (in two dimension $2/\pi$) is
reduced, as well. But still, the filamentary alignment quantified with
$\CF(r)$ indicates a strong alignment of the orientation of the
clusters with the connecting vector
(Fig.~\ref{fig:clusters-500-proj-orient}). Although the amplitude is
slightly reduced, this alignment is still visible out to scales of
100~\hMpc, as in the three dimensional analysis.

The question arises, why do we find two different regimes of the
vector correlations, using either $\CA(r)$ or $\CF(r)$. The direct
alignment, quantified with $\CA(r)$ extends out to 30~\hMpc, whereas
the filamentary alignment ($\CF(r)$) is visible out to 100~\hMpc.  A
glimpse on the real density distribution in
Fig.~\ref{fig:densityfield} sheds light on this topic. Clearly one can
see filamentary structures with sizes extending up to $100~\hMpc$, for
example the two pronounced filaments which end up in the knot at
$x\approx275~\hMpc$ and $y\approx75~\hMpc$.
Generically, such filaments are not straight but crinkled and show a
lot of small branches. This implies that the coherence of the angles
between the major axes of clusters harbored in these filamentary
structures is lost soon. Consequently the positive correlation seen
with $\CA(r)$ is confined to comparatively small scales
$\lesssim30~\hMpc$s (in the case of projected data even to
$\lesssim10~\hMpc$).  On the other hand, the signal exploring the
angle between orientation and connecting vector is not affected by the
small scale disorder.  Thus $\CF(r)$ is tracing the filamentary
alignment out to scales of 100~\hMpc.
\vspace*{0.3cm}
\section{Angular momentum, mass and spin correlations}
\label{sect:angular}
In a simulation we have access to the position {\em and} the velocity
of the particles. Hence, it is easy to determine the angular momenta
of our cluster sized halos.  In this section we use the direction of
the angular momenta of galaxy clusters as vector marks and compare
their correlation properties with the alignment seen in the
orientation of the mass distribution (Sect.~\ref{sect:shape}). With
scalar MCFs, using the absolute value of the angular momentum, the
cluster mass and the spin parameter as scalar mark, we will supplement
the foregoing result. 

\begin{figure}
\begin{center}
\resizebox{0.7\hsize}{!}{\includegraphics{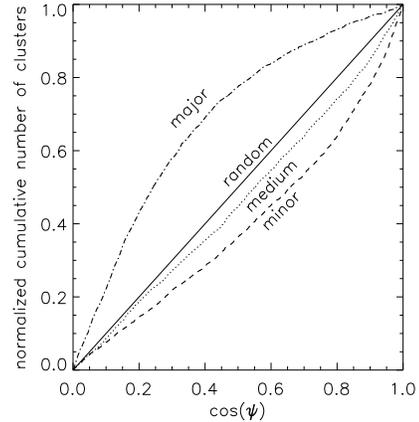}}
\end{center}
\caption{Cumulative plot of the number of clusters and the {\em cosine}
between the direction of the angular momentum and the major, medium
and minor cluster axis, respectively.  \label{fig:axan_kol}} 
\end{figure}

\begin{figure}
\begin{center}
\resizebox{0.7\hsize}{!}{\includegraphics{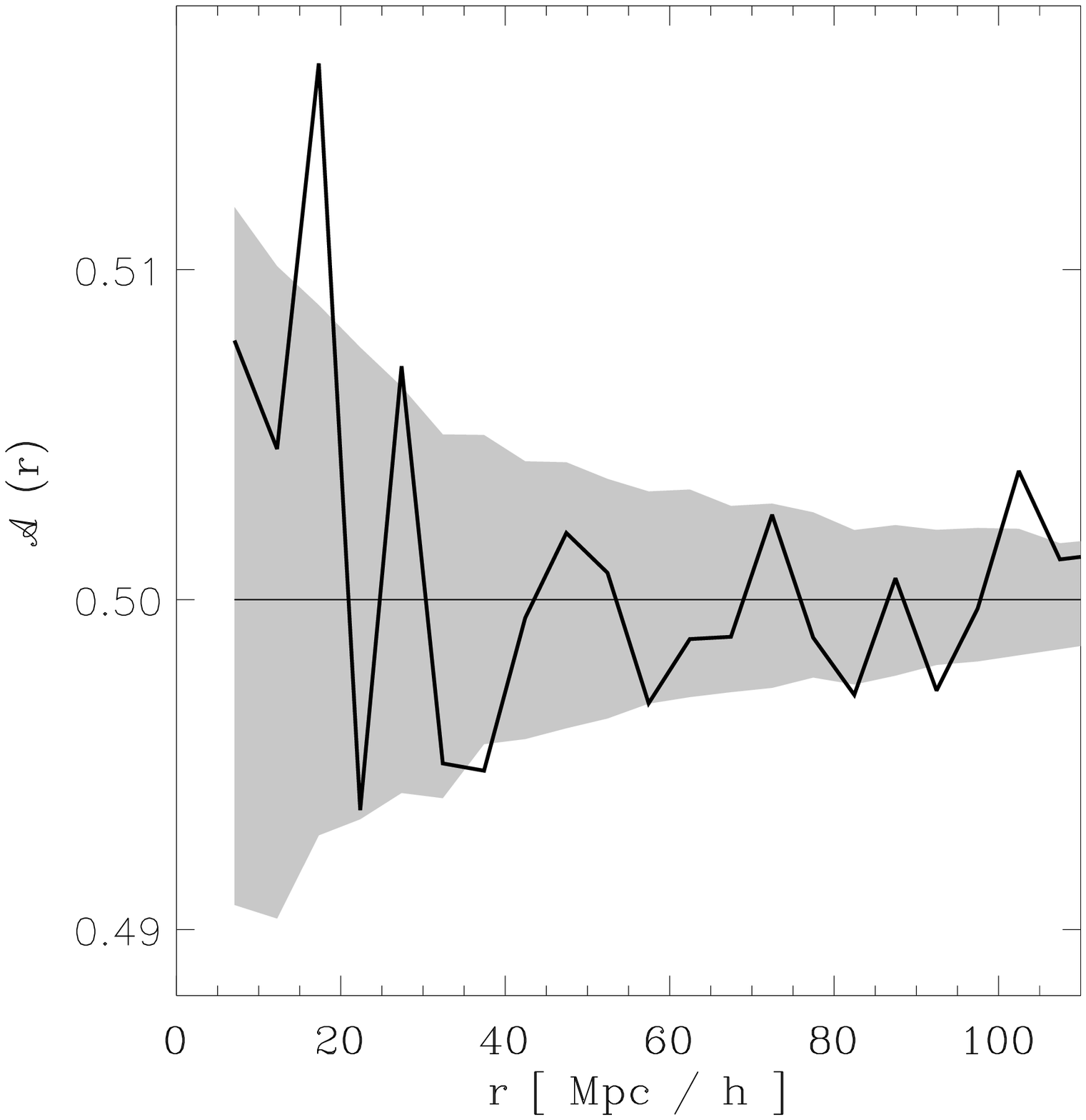}}
\resizebox{0.7\hsize}{!}{\includegraphics{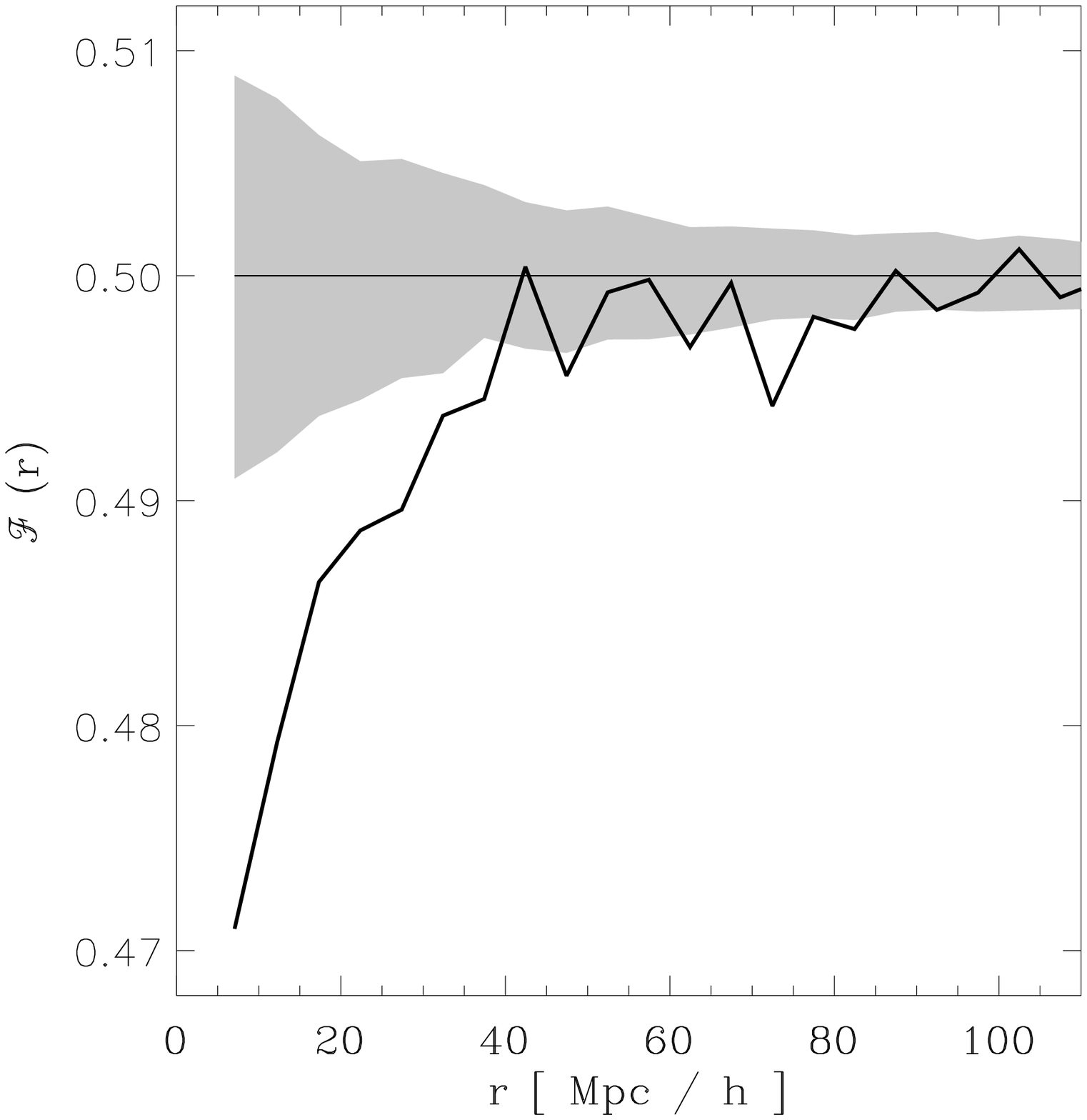}}
\end{center}
\caption{Mark correlation functions using  the normalized angular
  momentum $\bl$ with $|\bl|=1$ as vector mark. The shaded area is
  obtained by randomizing the orientation among the
  clusters.\label{fig:vec-angular-500}} 
\end{figure}

It is well known that there exists a correlation between the axes of
the mass ellipsoid and the direction of the angular momentum in
gravitationally bound N-body systems {}\citep{binney:galactic}.
Fig.~\ref{fig:axan_kol} shows that the major axis of the mass
ellipsoid tends to be perpendicular to the angular momentum. Compared
to the expectation from a purely random distribution, the minor axis
favors smaller angles with the angular momentum.
Hence, using the direction of the angular momentum as vector mark, we
are expecting to see effects comparable to the shape orientations.
The vector correlations of the directions of the angular momentum are
shown in Fig.~\ref{fig:vec-angular-500}.  There is no clear signal for
a direct alignment of the angular momenta as traced by $\CA(r)$.
However, the correlation $\CF(r)<0.5$ is smaller than for random
directions (shaded area), this indicates that the angular momenta are
preferably oriented perpendicular to the connecting vector $\hat\br$
on scales up to $30~\hMpc$.  The amplitude of the deviation
from the random orientation reaches approximately 5\% and is clearly
outside the fluctuations.
With Fig.~\ref{fig:axan_kol} in mind,
the latter result was expected from the correlations of the
orientations shown in Fig.~\ref{fig:clusters-500-orientation}: For a
pair of clusters residing in a filamentary structure the connecting
vector is oriented along this filament. The major axes of the
orientations tend to align with the filaments, and therefore the
angular momenta tend to stand perpendicular to the filaments.
A random orientation of the angular momenta in the planes
perpendicular to the filaments provides a simple explanation for the
absence of any signal in $\CA(r)$, still compatible with the strong
signal in $\CF(r)$.

\begin{figure}
\begin{center}
\resizebox{0.7\hsize}{!}{\includegraphics{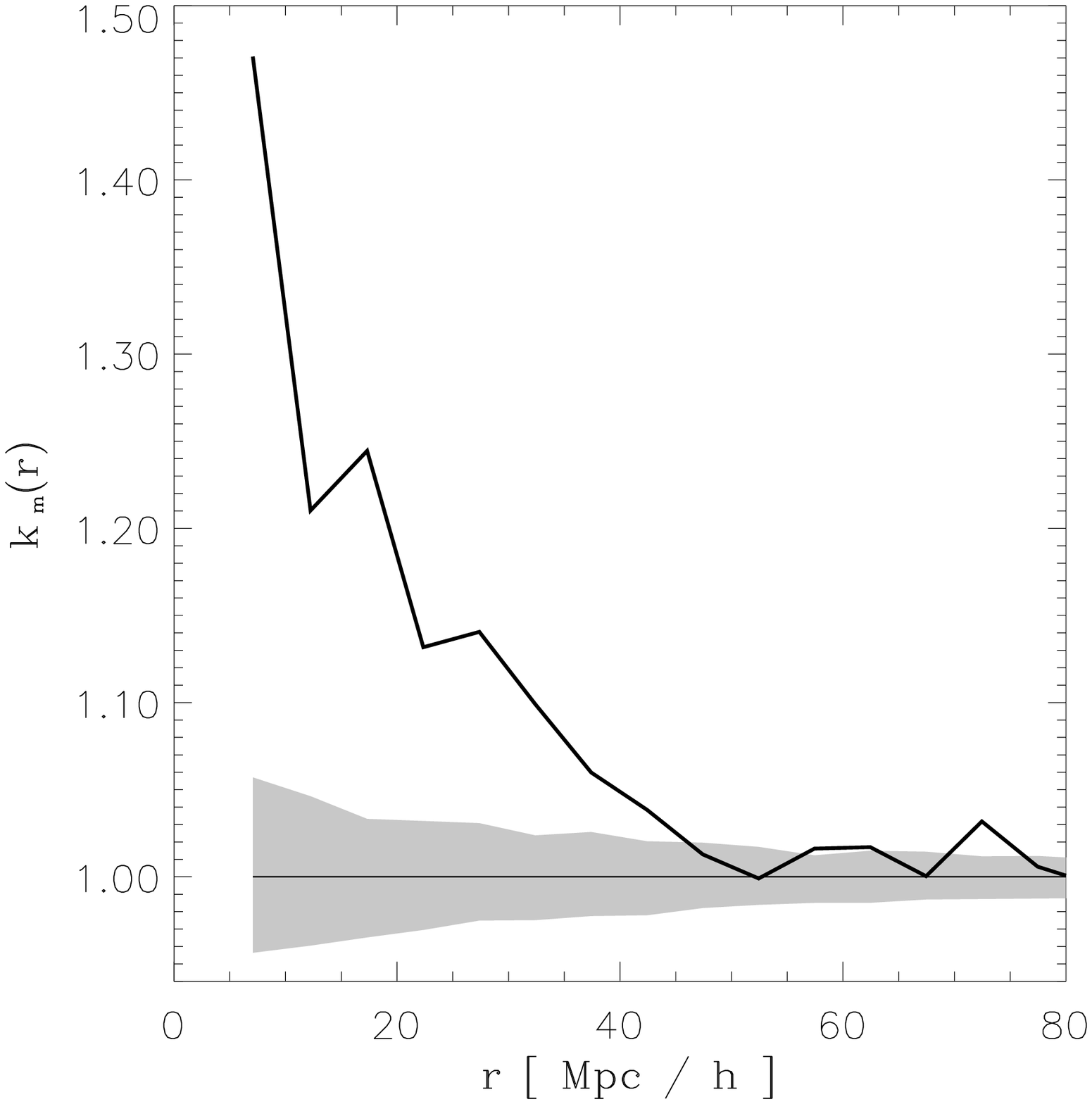}}
\resizebox{0.7\hsize}{!}{\includegraphics{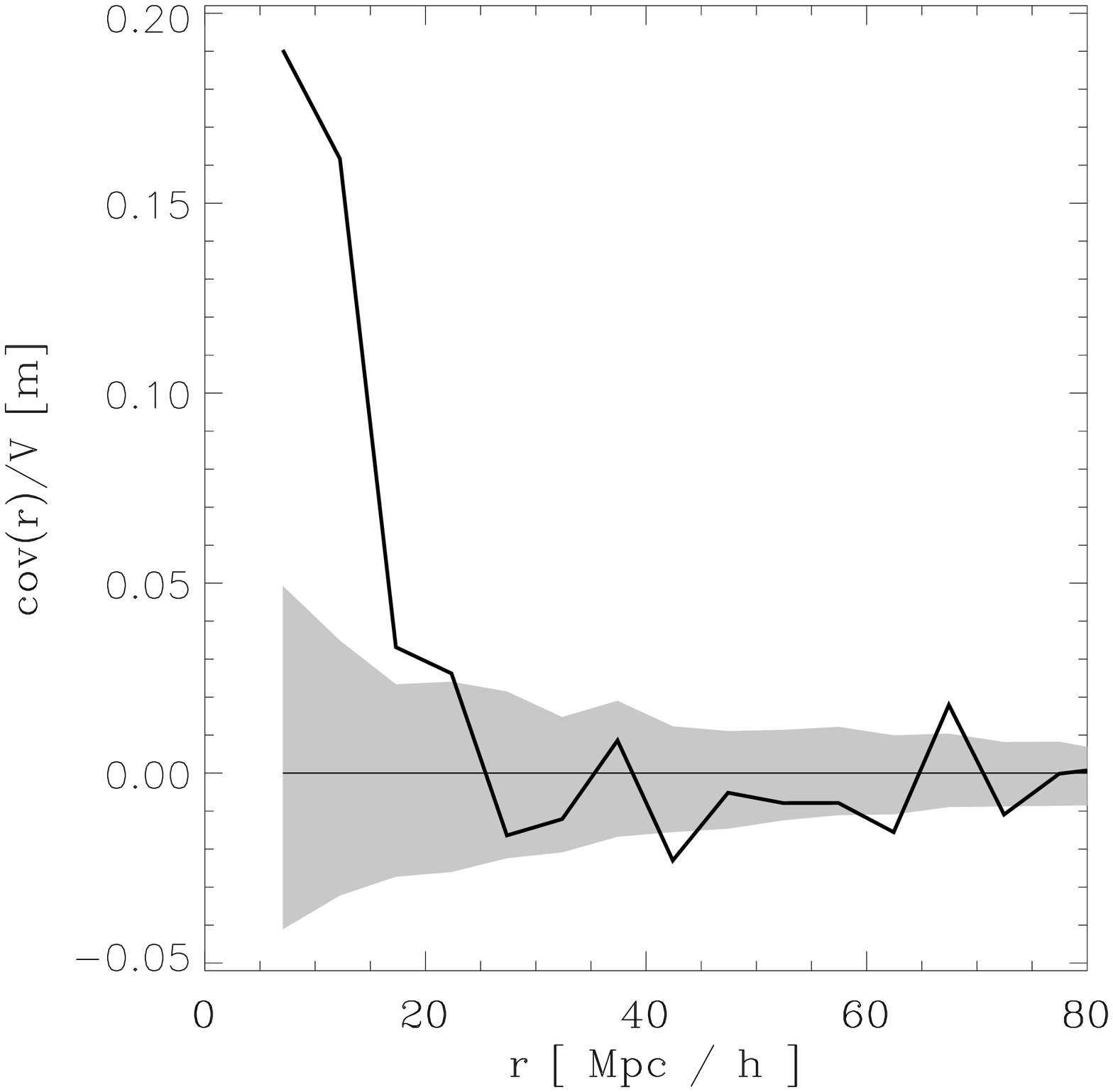}}
\end{center}
\caption{Mark correlation functions with the absolute value of the angular
momentum of the cluster as scalar mark. The shaded area is obtained
by randomizing the mark among the clusters.
\label{fig:scalar-moment-500}}
\end{figure}

Besides the correlations in the directions of the angular momentum
vectors, we are also interested in the correlations of their
magnitudes. We investigate these correlations with the MCFs discussed
in Sect.~\ref{sect:mcf-scalar}, using the magnitude of the angular
momentum as a scalar mark.  In Fig.~\ref{fig:scalar-moment-500} the
increased $k_{\rm m}(r)$ shows that pairs of clusters with separations
$\lesssim50~\hMpc$ tend to have higher mean angular momentum compared
to the overall mean angular momentum. The positive covariance up to
$\lesssim15~\hMpc$ indicates that both members of close pairs tend to have
similar angular momentum. Hence, inspecting both $k_{\rm m}$ and the
covariance, we see that close clusters tend to carry a similar amount
of angular momentum, larger than the overall mean angular momentum.

\begin{figure}
\begin{center}
\resizebox{0.7\hsize}{!}{\includegraphics{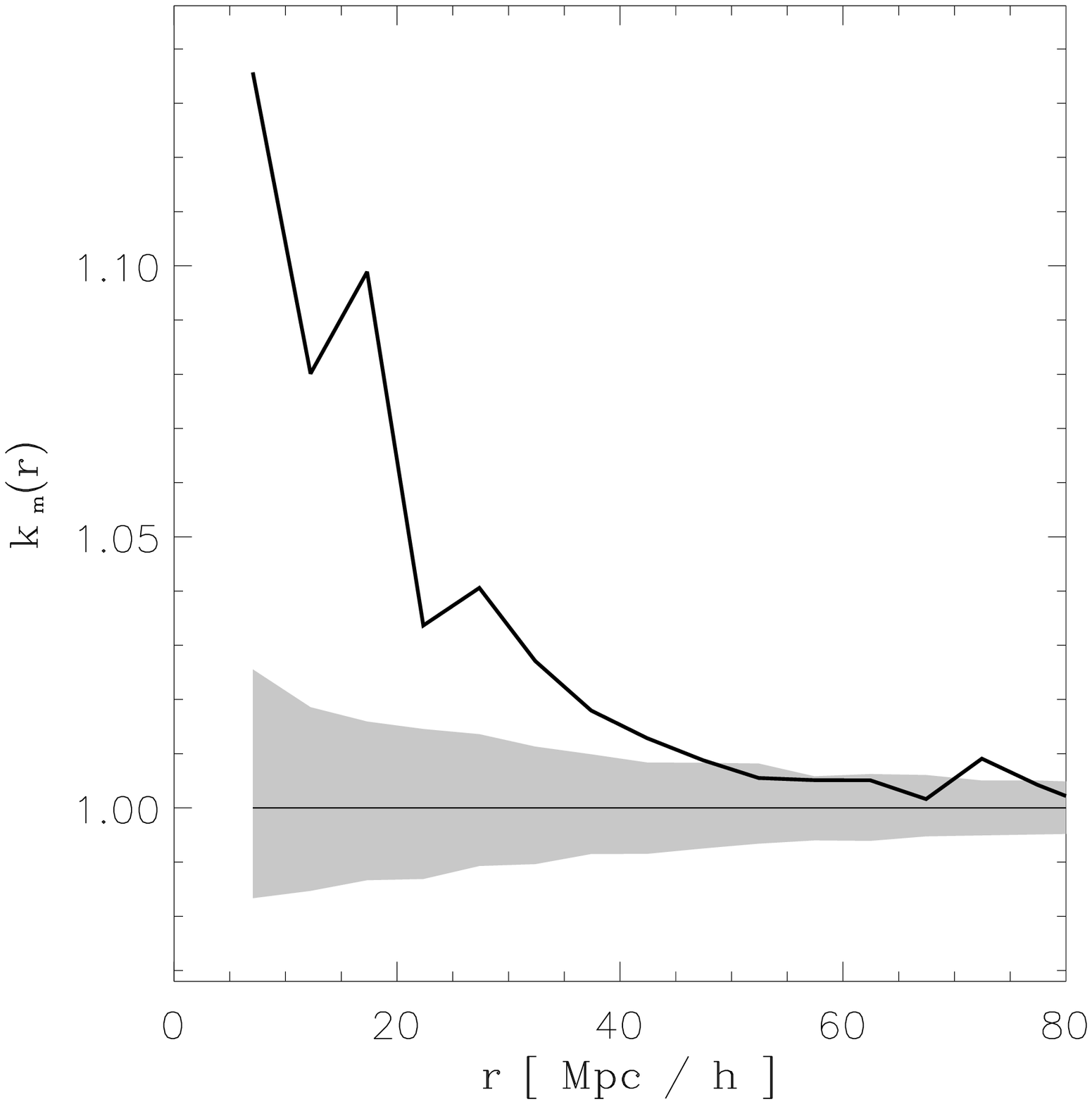}}
\resizebox{0.7\hsize}{!}{\includegraphics{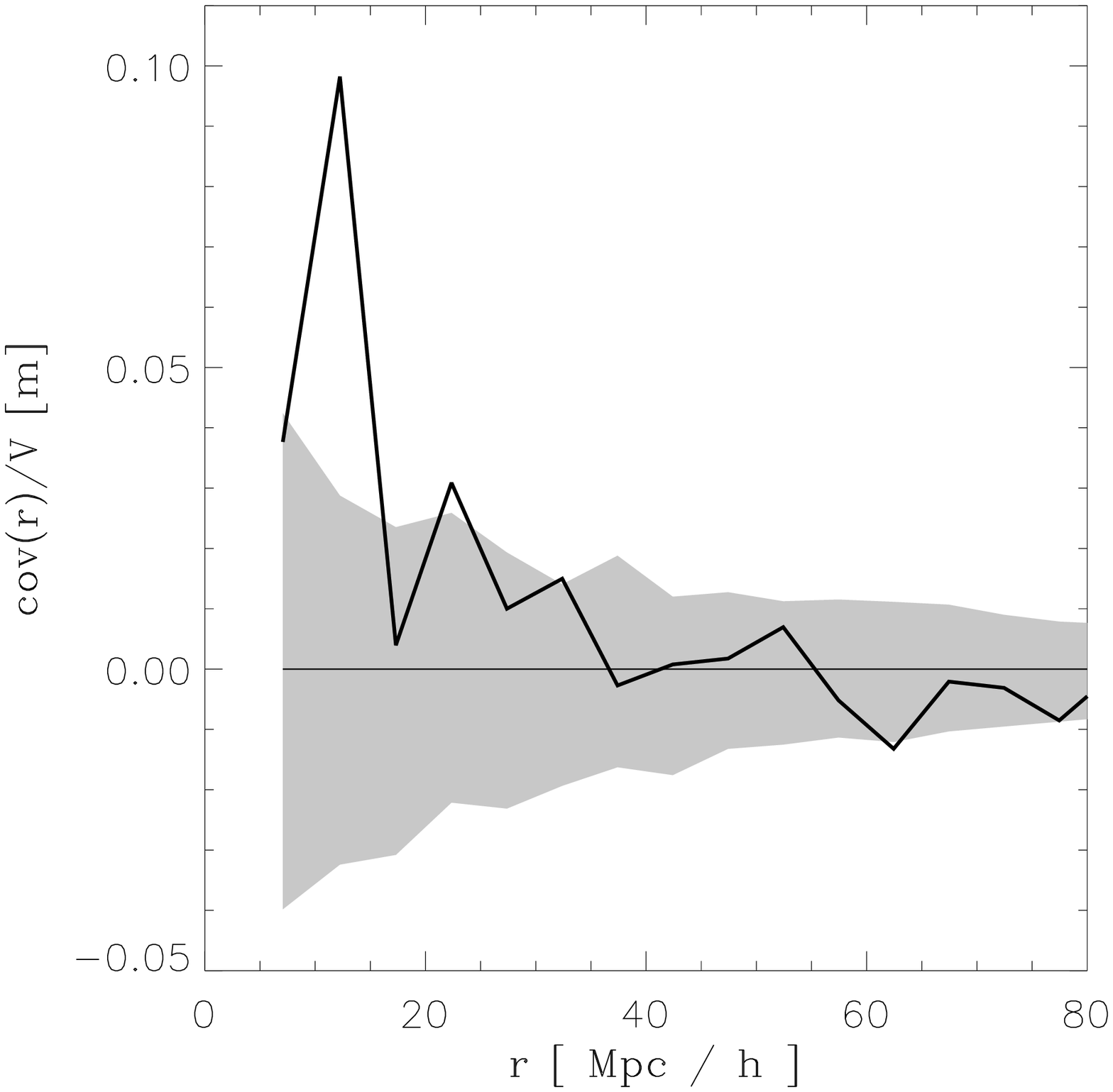}}
\end{center}
\caption{Mark correlation functions with the mass of the cluster halo
as scalar mark. The shaded area is obtained by randomizing the mark
among the cluster halos. \label{fig:mass-500}}
\end{figure}

\begin{figure}
\begin{center}
\resizebox{0.7\hsize}{!}{\includegraphics{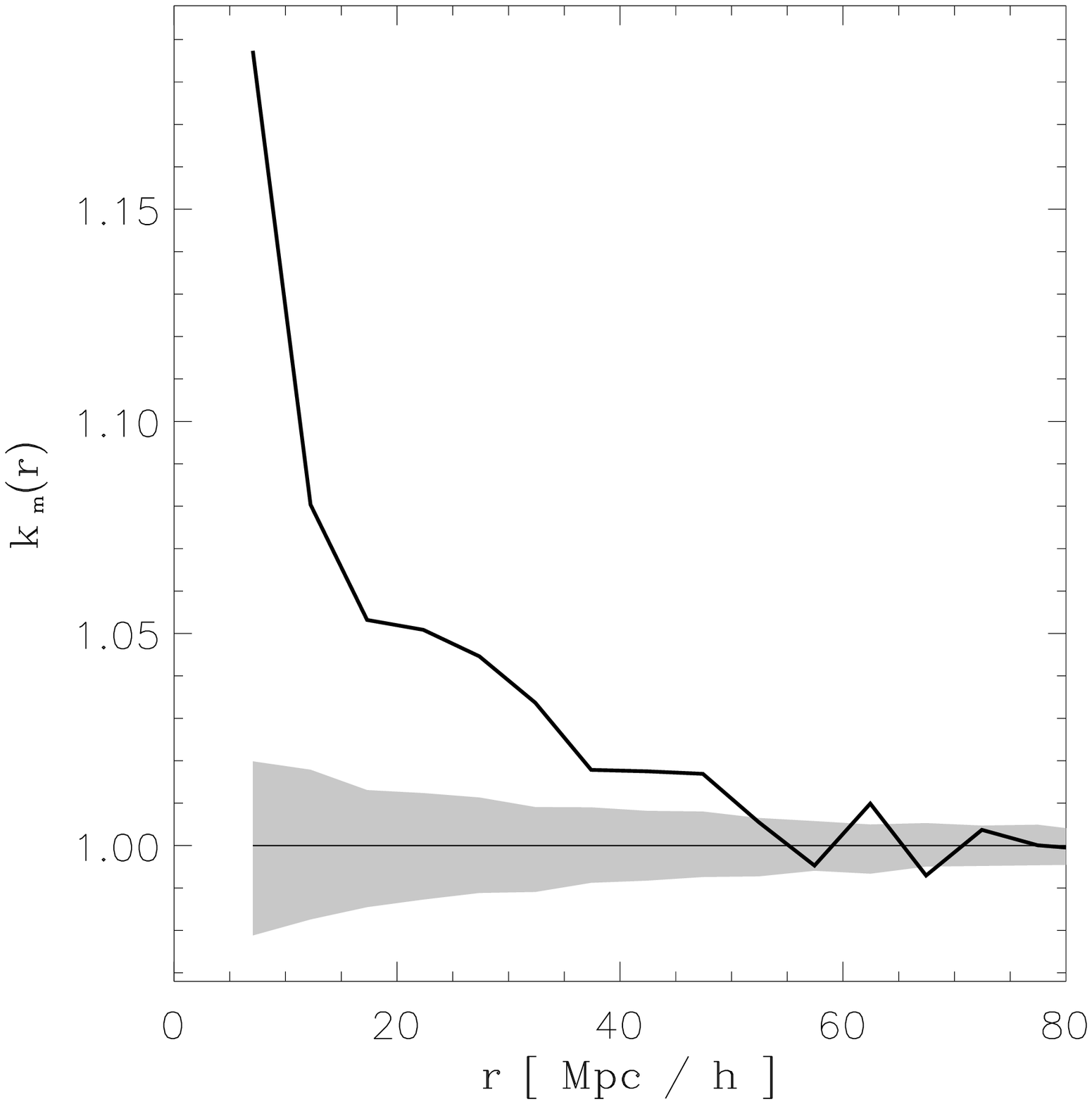}}
\resizebox{0.7\hsize}{!}{\includegraphics{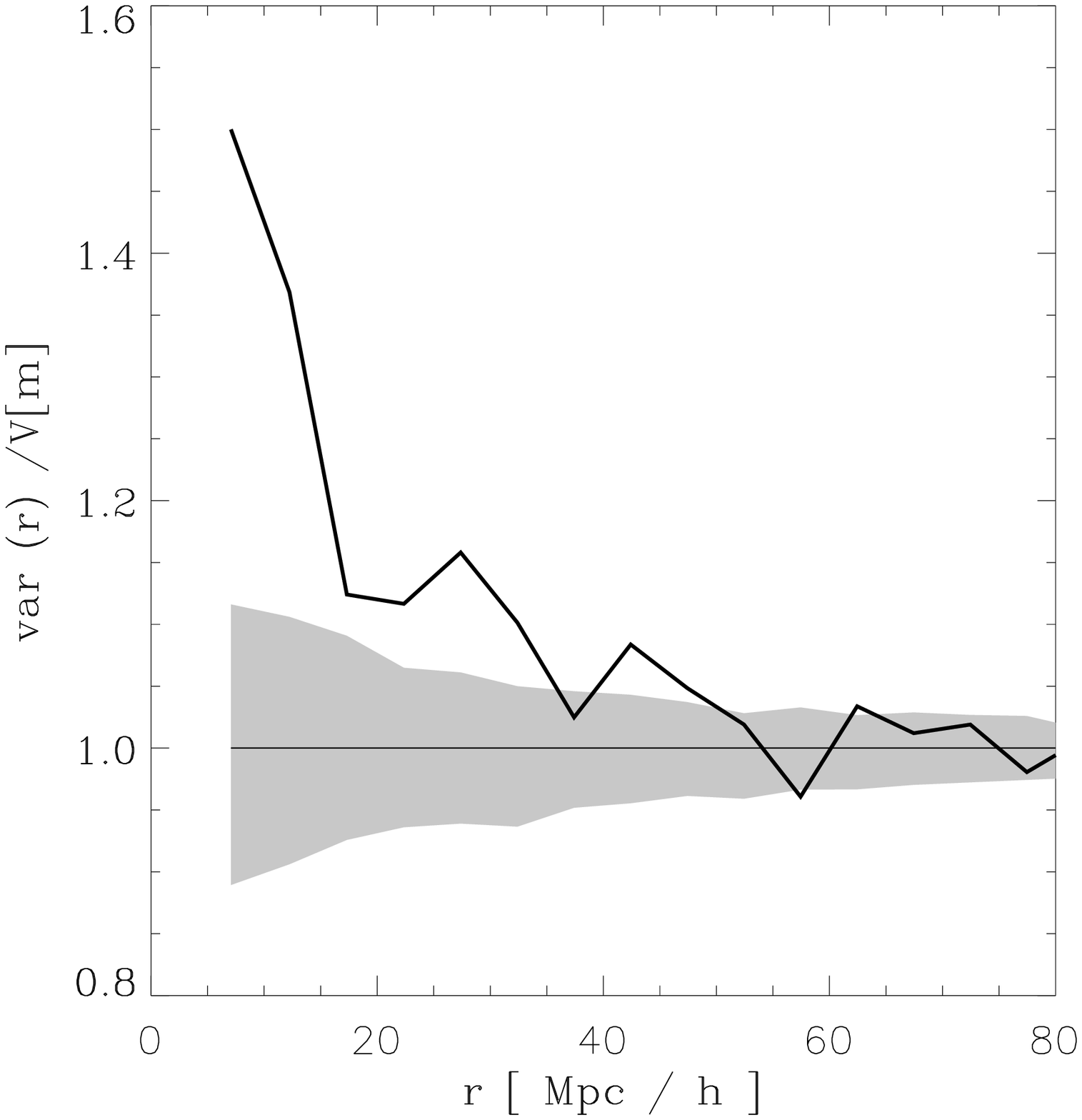}}
\end{center}
\caption{Mark correlation function with the spin parameter $\lambda$ of the 
cluster halo as scalar mark.  The shaded area is obtained by randomizing 
the mark among the clusters.
\label{fig:lambda-500}}
\end{figure}

There are two possibilities to explain this behavior. On the one hand
the mean mass of a close cluster pair could be enhanced, and so the
absolute value of the angular momentum would grow simply due to the
fact that bigger clusters are under consideration.  Indeed,
{}\citet{gottloeber:spatial} found such an increase in the mean mass
in close pairs (in their case traced by the maximum circular velocity
of the halo).  Consistently {}\citet{beisbart:luminosity} report an
enhanced luminosity for close galaxy pairs.  On the other hand, the
rotational support of close cluster pairs could be higher, meaning
that the spin parameter $\lambda$ is enhanced, cp.\
Eq.~(\ref{equ:spin}).  To illustrate this further, we investigate the
MCFs with the mass and the spin parameter as marks:
\newline
{\em Mass:} In Fig.~\ref{fig:mass-500} we show the MCFs of the cluster
distribution with the total mass as scalar mark.  The increased
$k_{\rm m}(r)$ indicates that close pairs of clusters tend to have higher
mean masses than the overall mean mass $\overline{m}$ on scales out to
50~\hMpc. The signal shows a deviation up to 10\% from a purely random
distribution. The conditional covariance $\cov(r)$ shows only a weak
positive signal, confined to scales below 15~\hMpc, indicating that
only close pairs tend to have similar masses.
\newline
{\em Spin:} In Fig.~\ref{fig:lambda-500} we show the MCFs of the
cluster distribution with the spin parameter $\lambda$ as scalar mark.
The increased $k_{\rm m}(r)$ indicates that neighboring pairs of clusters
tend to have higher spin parameter $\lambda$ compared to the overall
mean spin $\overline{\lambda}$. The signal seen in
Fig.~\ref{fig:lambda-500} shows an enlarged spin parameter of clusters
on scales out to 50~\hMpc, the same range as for the mass. The
deviation from random distribution is $\sim 15\%$.
Hence, not only an increased mass accretion but also tidal
interactions out to fairly large scales seem to characterize the
regions around cluster halos. The conditional covariance $\cov(r)$
shows no signal. This can be explained by looking at the mark variance
$\var(r)$. The signal at small distances deviates about $50\%$ from
random, indicating large scattering of the spin parameter. This large
dispersion also might be the reason why previous studies did not find
an environmental dependence of the amount of the spin parameter.  A
sufficiently large simulation and also a large number of clusters are
needed to get a clear signal in the MCFs for the spin parameter. 

Putting together all the results shown in
Figs.~\ref{fig:scalar-moment-500}\,-{}\ref{fig:lambda-500}, it turns
out that the increased mean angular momentum is not only caused by an
enhanced mass, but also by an enlarged spin parameter of neighboring
pairs of clusters.
From the present analysis we can not draw any firm conclusions,
whether this phenomenon is dominated by tidal interaction or
merging processes.  There are compelling observational hints (see
e.g.\ {}\citealt{plionis:substructure,schuecker:substructure}) that
clusters in high-density environments show indications of dynamical
disruption. Thus it seems to be likely that the enhancement of the
angular momentum of close neighbors is caused by a substantial mass
accretion (i.e.\ merging).  The concordant behavior of the mass and
the spin MCF support this interpretation, at least on scales below
$\sim15~\hMpc$. However, tidal interactions could be the cause for the
correlations seen on large scales. 

\section{Summary}
Whether there exist correlations in the orientations of galaxies or
galaxy clusters has been discussed for a long time.
{}\citet{binggeli:shape} reported a significant alignment of the
observed galaxy clusters out to
50~\hMpc. {}\citet{struble:new,struble:new-erratum} claimed that this
effect is small and prone to systematics and {}\citet{ulmer:major}
find no indication in their investigation.  Subsequently, several
authors found, sometimes only weak, signs of alignments in the galaxy
and galaxy cluster distribution (see e.g.\
{}\citealt{djorgovski:coherent,
lambas:statistics,fuller:alignments,heavens:intrinsic}).
As a novel statistical method we have used the mark correlation
functions (MCFs) to quantify the alignment of cluster sized halos,
extracted from a large scale simulation based on a $\Lambda$CDM
cosmology. Our sample with 3000 cluster sized halos is bigger than the
currently available samples of galaxy clusters. The unambiguous signal
we obtain benefits from the large statistics in our simulation.

Using two different weighting functions in the construction of the
MCFs we investigate the direct alignment and the filamentary
alignment. First we use the major axis of the mass ellipsoid as our
direction marker. The clear signal from the direct alignment $\CA(r)$
extends out $\sim30~\hMpc$. For the filamentary alignment $\CF(r)$ we
find deviations from isotropy up to $\sim100~\hMpc$.  Considering the
projected mass distribution, the signal from the direct alignment $\CA(r)$ 
already vanishes at a scale of $\sim10~\hMpc$. However, we find a
filamentary alignment $\CF(r)$ out to scales of $\sim100~\hMpc$, even for 
the projected data.  This scale is very similar to the size of the large
scale filaments seen in our simulation. We think that the function  
$\CF(r)$ is a powerful tool for exploring large scale alignment effects 
also in observational data.

{}\citet{franx:elliptical} showed that the angular momentum of an
ellipsoidal system tends to align with the minor axis of this
system. We confirm this behavior in our simulation.  With the
angular momentum as vector mark, $\CF(r)$ shows the expected
filamentary correlations: the angular momentum tends to be
perpendicular to the connecting line, i.e.\ the filament, up to
separations of $\sim40~\hMpc$.  However, we obtain no signal for
the direct alignment $\CA(r)$. This is in concordance with the perception 
that the angular momenta are randomly oriented in the planes perpendicular 
to the filaments.

With the scalar MCFs $k_{\rm m}(r)$ and $\cov(r)$ we have investigated the
correlations in the absolute value of the angular momentum.  Close
pairs of clusters tend to have similar and also higher absolute values
of the angular momentum compared to the global average. A clear signal
can be detected up to $\sim50~\hMpc$.
A further analysis of the mass and spin parameter distribution of the
clusters with the MCFs has shown that this enhancement of the absolute
value of the angular momentum is caused by an enhanced mass of close
pairs of clusters as well as by the stronger rotational support of 
them. 
This behavior should be caused by the combined action of large-scale
tidal fields and the hierarchical merging of progenitor structures and
mass inflow onto the cluster. Since this mass growth follows the large
scale filaments, tidal interactions and merger events are tightly
connected. The mark correlation function with scalar and vector marks
deliver quantitative measures of these effects. 

\subsection*{Acknowledgments}
M.K. was supported by the Sonder\-for\-schungs\-bereich 375 f\"ur 
Astro--Teilchen\-physik der Deutschen For\-schungs\-ge\-mein\-schaft.

\bibliographystyle{aa}

\end{document}